\documentclass[12pt]{article}
\pdfoutput=1 

\usepackage[a4paper, margin=0.8in]{geometry}
\usepackage
 {
   amsmath,amssymb,graphicx,tabularx
 }
 
\usepackage[usenames,dvipsnames]{xcolor}
 
\definecolor{X575}{rgb}{0.05, 0.7, 0.05}

\usepackage{caption} 
\usepackage{color}
\usepackage{hyperref}
\usepackage{graphicx}
\usepackage{booktabs}
\usepackage{multirow}
  \usepackage{bm}
  \usepackage{rotating}
  \usepackage{floatpag}
  \usepackage{mathrsfs}
  \rotfloatpagestyle{empty}

  \usepackage{amssymb}
  \usepackage{amsthm}
  \usepackage{graphicx}
  \usepackage{multind}

\title{\bf What really matters  in Hilbert-space stochastic processes}

\begin{document}
\maketitle
\flushbottom

\begin{center}
Giancarlo Ghirardi$^a$, Oreste Nicrosini$^b$ and Alberto Rimini$^c$ 
\end{center}

\begin{center}
\it 
$^a$Abdus Salam ICTP, Strada Costiera 11, 34151 Trieste, Italy \\
$^b$Istituto Nazionale Fisica Nucleare, Sezione di Pavia \\ 
via A. Bassi 6, 27100 Pavia, Italy \\
$^c$Dipartimento di Fisica, Universit\`a di Pavia \\ 
via A. Bassi 6, 27100 Pavia, Italy
\end{center}

%

\abstract{The relationship between discontinuous and continuous stochastic processes in Hilbert space is investigated. 
It is shown that for any continuos process there is a parent discontinuous process, that becomes the continuous one 
in the proper  infinite frequency limit. From the point of view of solving the quantum  measurement 
problem, what really matters is the choice of the set of operators whose value distributions are made sharp. In particular, 
the key role of position sharping is emphasized. }


%
\def\ket#1{\vert #1 \rangle}
\def\bra#1{\langle #1 \vert}
\def\braket#1#2{\langle #1 | #2 \rangle}
\def\diff{{\rm d}} 
\font\tenbi=cmmib10 
\def\myfrac#1#2{{#1\over{#2}}} 
\def\norm#1{\Vert #1 \Vert}
\font\tenrsfs=rsfs10

\def\buildler#1\over#2{\mathrel{\mathop{#1}\limits_{#2}}}
\def\tofor#1{\buildler \longrightarrow \over {#1}}
\def\ttofor#1{\buildler \llongrightarrow \over {#1}}

\def\llongrightarrow{\relbar\mkern-10mu{\relbar\mkern-10mu
{\relbar\mkern-10mu{\longrightarrow}}}}

\def\compint{\!\! \int \!\!}
\def\mumb{- {\textstyle \myfrac{1}{2}} \beta}

\section{Introduction}

In attempting to overcome the difficulties connected with the 
quantum measurement problem, two types of stochastic processes in Hilbert space have been considered, 
the so-called hitting or discontinuous processes and the continuous ones. All  proposed processes aim at 
describing reduction as a physical process. A common feature is non-linearity, which originates from the need 
of maintaining normalization while reducing the state vector. One finds sometimes the statement 
that the continuous processes allow to deal with systems which cannot be treated with the hitting processes. 
The aim of the present contribution is to discuss the relationship between discontinuous and continuous processes. 
 Actually, it will be shown that all continuous processes 
have a physically equivalent {\it parent} discontinuous one, applicable to the same class of physical systems. 

In all examples of stochastic process in  Hilbert space a set of {\it physical quantities} (observables) appears, 
represented by  the corresponding set of selfadjoint operators. 
The  process acts inducing the {\it sharpening } of the  distribution of values of those quantities around a 
stochastically chosen centre. What really matters from the physical point of view is the choice of the set 
of quantities, rather than the details of the sharpening procedure. 

Both approaches have their benefits. In the discontinuous processes the  meaning of the parameters 
appearing in their definition is 
 transparent and the physical consequences are easy to grasp, 
while the continuous processes are undoubtedly more elegant from the mathematical point of view. 
The continuous processes depend on  one parameter less than the corresponding 
discontinuous ones. 
But this difference is  illusory since, as discussed below,  the effectiveness of the discontinuous process 
depends on  two of its parameters  only through their  product. 

In the next section a simplified heuristic derivation of a generic continuous process as the proper infinite frequency limit of 
the parent hitting process is given. In the last section the three physically most relevant examples of discontinuous and corresponding continuous 
processes are presented. 

\section{Infinite frequency limit}
\label{sect:ifl}

In the present section we show how a generic hitting process becomes a continuous stochastic  process in the Hilbert space 
under a suitable infinite frequency limit. 

A hitting process is characterized by a choice of the quantities whose distribution is to be made sharp, 
by the mean frequency of 
the Poisson distribution 
of hitting times, and by  the accuracy of the sharpening of the chosen quantities. 
A probabilistic rule is assumed for the distribution of the hitting centres. 
The effectiveness of the hitting process actually depends on the products of the mean frequency times the accuracy parameters, so that 
a given effectiveness can be obtained by increasing the hitting frequency and, at the same time, appropriately decreasing the accuracy parameters. 
Taking the infinite frequency limit with this prescription, one gets the corresponding continuous process. 

The above feature of the effectiveness of a hitting process was first noticed by~\cite{Barchielli:1982ft} in a different conceptual context. Later, the same feature was highlighted by~\cite{Ghirardi:1985mt} and the infinite frequency limit was considered by~\cite{Ghirardi:1989cn} with reference to the time evolution of the statistical operator. Last,~\cite{Diosi:1988tf} in a particular case 
and~\cite{NicrosiniRimini:1990} in the general case considered the same limit for the time evolution of the state vector, obtaining the continuous stochastic process.  

In what follows for the sake of simplicity we use evenly spaced hitting times instead of random ones, and, at least initially, we ignore the Schr\"odinger evolution. 
In the present section the sharpened quantities are assumed, for simplicity, to have a purely discrete 
spectrum.

Let the set of compatible quantities characterizing the considered discontinuous stochastic process be 
\begin{equation}
{\hat{\bm A}} \equiv \{ \hat A_p;p=1,2,\dots,K\}, \qquad  \big[\hat A_p\,,\!\hat A_q\big]=0, \qquad \hat{A}^\dagger_p = \hat{A}_p , 
\end{equation}
and the sharpening action be given by the operator
\begin{equation}
S_i = \left(\myfrac{\beta}{\pi}\right)^{\!K\!/\!4} \! \exp(R_i) , \qquad R_i= \mumb \,(\hat{\bm A} -\bm a_i)^2 . 
\end{equation}
The parameter $ \beta $ rules the accuracy of the sharpening and $\bm a_i $ is the centre of the hitting $i$. 
It is assumed that the hittings occur with frequency $\mu$. 

The sharpening operator for the $i$-th hitting $S_i$ acts on the normalized state vector $\ket{\psi_t}$ giving the normalized state vector $\ket{\psi'_{i,t}}$ according to 
\begin{equation}
\label{singlehitting}
\ket{\psi'_{i,t}} = \myfrac{\ket{\chi_{i,t}}}{\norm{\chi_{i,t}}} , 
\hskip50pt
\ket{\chi_{i,t}} = S_i \ket{\psi_{t}} .  
\end{equation}
The probability that the hitting takes place around $\bm a_i $ is 
 \begin{equation}
 \label{singleprob}
{\mathscr P} \left( \psi_t \vert \bm a_i  \right) = {\norm{\chi_{i,t}}}^2 .
\end{equation}
When considering $n$ evenly spaced hittings in $(t,t+\diff t]$, so that the time interval between two adjacent hittings is 
$\tau=\diff t/n$, the hitting frequency is given by $\mu=1/\tau$. In this case, assumptions (\ref{singlehitting}) and (\ref{singleprob}) become 
 \begin{equation}
 \label{multihitting}
\ket{\psi_{t+i\tau}} = \myfrac{\ket{\chi_{t+i\tau}}}{\norm{\chi_{t+i\tau}}} , 
\hskip50pt
\ket{\chi_{t+i\tau}} = S_i \ket{\psi_{t+(i-1)\tau}} , 
\end{equation}
with probability 
\begin{equation}
 \label{multiprob}
 {\mathscr P}(\psi_{t+(i-1)\tau}|\bm a_i) = {\mathscr P}(\bm a_1,\bm a_2,\dots,\bm a_{i-1}| \bm a_i)
= \norm{\chi_{t+i\tau}}^2 . 
\end{equation}

The continuous process based on the same quantities ${\hat{\bm A}} \equiv \{ \hat A_p;p=1,2,\dots,K\}$ is ruled by 
the It{\^ o} stochastic differential equation 
\begin{equation}
\diff\ket{\psi} 
\label{contproc}
= \Big[\sqrt{\gamma}\big(\hat{\bm A} \!-\!\langle\hat{\bm A}\rangle_{\!\psi_t}\big) \cdot \diff\bm B
-{\textstyle\myfrac{1}{2}} \gamma \big(\hat{\bm A} \!-\!\langle\hat{\bm A}\rangle_{\!\psi_t}\big)^2 \diff t \Big]
\,\ket\psi , 
\end{equation}
where
\begin{equation}
\langle\hat{\bm A}\rangle_{\!\psi_t} = \bra{\psi_t } \hat{\bm A} \ket{\psi_t } 
\end{equation}
and 
\begin{equation}
\diff{\bm B}\equiv\{\diff B_p;p=1,2,\dots,K\}, \qquad \overline{\diff\bm B} = 0 , \qquad \overline{\diff B_p\,\diff B_q} = 
\delta_{pq} \,\diff t . 
\end{equation}
The parameter $\gamma$ sets the effectiveness of the process. 

We shall show that taking the infinite frequency limit of the discontinuous process (\ref{multihitting}) and (\ref{multiprob}) with the prescription 
\begin{equation}
\label{eq:betamu}
\beta \mu = {\rm constant} = 2 \gamma , 
\end{equation}
one gets the continuous process (\ref{contproc}). As a consequence it becomes apparent that, for $t \to \infty$, the continuous process drives the state vector to a common eigenvector of the operators $\hat {\bm A}$, 
the probability of a particular eigenvector $\ket{\bm a_r}$ being 
$\vert \braket{\bm a_r}{\psi_0} \vert^2$, for the state vector $\ket{\psi_0}$  at a given arbitrary initial time. 

In what follows for any set of stochastic variables  $\bm v$ we use the notation $\overline{v_p}$  for the mean value and  
$\overline{\overline{v_p v_q}} = \overline{(v_p-\overline{v_p})(v_q-\overline{v_q})}$ for the variances and covariances. 

The effect of  $n$ hitting processes in the time interval $(t,t+\diff t]$ is described by 
\begin{equation}
\ket{\chi_{t+dt}} = \left( \myfrac{\beta}{\pi} \right)^{n K / 4} \exp(R_n) \dots  \exp(R_2) \exp(R_1) \,\ket{\psi_t}, 
\hskip 15pt 
\ket{\psi_{t+dt}} = \myfrac{\ket{\chi_{t+dt}}}{\norm{\chi_{t+dt}}} . 
\end{equation}
By using the properties of the exponential function, the final non-normalized state vector is then given by 
\begin{eqnarray}
\label{eq:statechi}
&&\ket{\chi_{t+dt}}  = \left( \myfrac{\beta}{\pi} \right)^{n K / 4} \! \! \! \! \exp \Big(\! \! \mumb \,{\sum_{i=1}^n} \bm a_i^2 \Big) 
\exp\Big\{\,{\sum_{i=1}^n} \Big[ \!\mumb
\Big(\hat{\bm A}^2 -2 \hat{\bm A}\cdot\bm a_i \Big)\Big]\Big\} \ket{\psi_t}  \nonumber \\
&&= F \exp \Big\{\!\mumb \Big[ n\,\hat{\bm A}^2 - 2\,\,n\,\hat{\bm A} \cdot {\textstyle\myfrac{1}{n} }
{\sum_{i=1}^n}  \bm a_i \Big] \Big\} \ket{\psi_t} \nonumber \\
&&= F \exp \Big\{\!\mumb\,n\Big[ \hat{\bm A}^2 
- 2\,\hat{\bm A }\cdot {\textstyle\myfrac{1}{n}} {\sum_{i=1}^n}  
\big(\bm a_i - \langle\hat{\bm A}\rangle_{\!\psi_t}\big)
- 2\,\hat{\bm A} \cdot \langle\hat{\bm A}\rangle_{\!\psi_t} 
\Big] \Big\} \ket{\psi_t} , 
\end{eqnarray}
with joint probability 
\begin{equation}
{\mathscr P} (\bm{a_1}, \cdots, \bm{a_n}) = \norm{\chi_{t+dt}}^2 . 
\end{equation}
Actually, in the limit $\beta \to 0$ the joint probability factorizes as 
\begin{equation}
{\mathscr P} (\bm{a_1}, \cdots, \bm{a_n}) = {\mathscr P} (\bm{a_n}) \cdots {\mathscr P} (\bm{a_1}) 
\end{equation}
where 
\begin{equation} 
{\mathscr P}(\bm a_i) = \Big(\myfrac{\beta}{\pi}\Big)^{\!K\!/\!2} \! 
\bra{\psi_t} \exp\big(\!-\beta\,(\hat{\bm A} - \bm a_i)^2 \big) \ket{\psi_t} 
\end{equation}
as illustrated in the following two-hitting example. 
The joint probability for two hittings is given by 
\begin{equation}
{\mathscr P} (\bm{a_1}, \bm{a_2}) = {\mathscr P} (\bm{a_1} \vert \bm{a_2})  {\mathscr P} (\bm{a_1}) , 
\end{equation}
where ${\mathscr P} (\bm{a_1} \vert \bm{a_2})$ is the conditional probability of ${\bm a_2}$ given ${\bm a_1}$. In turn, 

\begin{eqnarray}
&&{\mathscr P}(\bm a_1|\bm a_2) = \norm{\chi_{t+2\tau}}^2 \qquad \qquad \nonumber  \\ 
&&= \Big(\myfrac{\beta}{\pi}\Big)^{\!K\!/\!2}
\!\bra{\psi_{t+\tau}} \,\!\!\exp\,(-\beta\,(\hat{\bm A} -\bm a_2)^2 \ket{\psi_{t+\tau}} \nonumber \\
&&= \Big(\myfrac{\beta}{\pi}\Big)^{\!K\!/\!2} \Bigg[ \Big(\myfrac{\beta}{\pi}\Big)^{\!-K\!/\!2} \! 
\bra{\psi_{t}} \exp\,(-\beta\,(\hat{\bm A} -\bm a_1)^2 \ket{\psi_{t}}^{\!-1}\Bigg]  \nonumber \\
&& 	\qquad \bra{\chi_{t+\tau}} \exp\,(-\beta\,(\hat{\bm A} -\bm a_2)^2 \ket{\chi_{t+\tau}} \nonumber \\
&&=  \Big(\myfrac{\beta}{\pi}\Big)^{\!K\!/\!2} \Bigg[ \Big(\myfrac{\beta}{\pi}\Big)^{\!-K\!/\!2} \! 
\bra{\psi_{t}} \exp\,(-\beta\,(\hat{\bm A} -\bm a_1)^2 \ket{\psi_{t}}^{\!-1}\Bigg] \nonumber \\ 
&& \qquad \Big(\myfrac{\beta}{\pi}\Big)^{\!K\!/\!2}
\bra{\psi_{t}} \exp\,(-\beta\,(\hat{\bm A} -\bm a_2)^2 
\exp\,(-\beta\,(\hat{\bm A} -\bm a_1)^2 \ket{\psi_{t}} \nonumber \\
&& \ttofor{\beta\rightarrow0} \Big(\myfrac{\beta}{\pi}\Big)^{\!K\!/\!2}
\bra{\psi_{t}} \exp\,(-\beta\,(\hat{ \bm A} -\bm a_2)^2  \ket{\psi_{t}} 
= {\mathscr P}(\bm a_2)
\end{eqnarray}
We now compute the statistical properties of the variables $\bm a_i$. The average value is given by 
\begin{eqnarray}
\overline{\bm a_i} ~\ttofor{\beta\rightarrow0} 
&&\compint\diff\bm a_i \, \bm a_i\,{\mathscr P}(\bm a_i) 
= \compint\diff\bm a \, \bm a\,{\mathscr P}(\bm a) \nonumber \\
&&= \compint\diff\bm a \, \bm a \,\Big(\myfrac{\beta}{\pi}\Big)^{\!K\!/\!2} \! 
\bra{\psi_t} \exp\big(\!-\beta\,(\hat{\bm A} - \bm a)^2 \big) \ket{\psi_t} . 
\end{eqnarray}
By inserting the expansion of the identity in terms of the common eigenvectors of the operators $\bm A$, satisfying   
$\hat A_p \,\ket{\bm\alpha_k} = \alpha_{kp} \ket{\bm\alpha_k}$,   one finds  
\begin{eqnarray}
\overline{\bm a_i} ~\ttofor{\beta\rightarrow0} &&= {\sum_k} \,\compint\diff\bm a \, \bm a \,\Big(\myfrac{\beta}{\pi}\Big)^{\!K\!/\!2} \! 
\bra{\psi_t} \exp\big(\!-\beta\,(\bm\alpha_k - \bm a)^2 \big) \ket{\bm\alpha_k}\braket{\bm\alpha_k}{\psi_t}   \nonumber \\
&&= {\sum_k} \big| \braket{\bm\alpha_k}{\psi_t} \big|^2
\compint\diff\bm a \, \bm a \,\Big(\myfrac{\beta}{\pi}\Big)^{\!K\!/\!2} \! 
\exp\big(\!-\beta\,(\bm\alpha_k - \bm a)^2 \big)  \nonumber \\
&&= {\sum_k} {\cal P}_{\!\psi_t}\!(\bm\alpha_k) \,\bm\alpha_k = \langle\hat{\bm A}\rangle_{\!\psi_t} . 
\end{eqnarray}
Similarly, for the variances  one gets
\begin{eqnarray}
&&\overline{\overline{a_{ip}^2}} ~\ttofor{\beta\rightarrow0} 
\compint\diff\bm a_i \, a_{ip}^2\,{\mathscr P}(\bm a_i)  -  \overline{a_{ip}}^2
= \compint\diff\bm a \, a_{p}^2\,{\mathscr P}(\bm a)  -  \overline{a_{p}}^2 \hphantom{aaaaaaaaaaaa} \nonumber \\
&&= \compint\diff a_p \, a_{p}^2 \,\Big(\myfrac{\beta}{\pi}\Big)^{\!1\!/\!2} \! 
\bra{\psi_t} \exp\big(\!-\beta\,(\hat A_p - a_p)^2 \big) \ket{\psi_t}  -  \overline{a_{p}}^2 \nonumber \\
&&= {\sum_k} \,\compint\diff a_p \, a_{p}^2 \,\Big(\myfrac{\beta}{\pi}\Big)^{\!1\!/\!2} \! 
\bra{\psi_t} \exp\big(\!-\beta\,(\alpha_{kp} - a_p)^2 \big) \ket{\tenbi\alpha_k}\braket{\tenbi\alpha_k}{\psi_t} 
-  \overline{a_{p}}^2 . 
\end{eqnarray}
By properly shifting the integration variable one then finds 
\begin{eqnarray}
\overline{\overline{a_{ip}^2}} ~\ttofor{\beta\rightarrow0} && {\sum_k} \big| \braket{\bm\alpha_k}{\psi_t} \big|^2
\compint\diff b \, (b+\alpha_{kp})^2 \,\Big(\myfrac{\beta}{\pi}\Big)^{\!1\!/\!2} \! 
\exp\big(\!-\beta\,b^2\big) -  \overline{a_{p}}^2 \nonumber \\
\cr\noalign{\vskip5pt} 
&&= {\sum_k} {\cal P}_{\!\psi_t}\!(\bm\alpha_k) \left( \myfrac{1}{2\beta} + \alpha_{kp}^2\right) 
-  \overline{a_{p}}^2  \nonumber \\
&& = \myfrac{1}{2\beta} +  \langle\hat A_p^2\rangle_{\!\psi_t} - \langle\hat A_p\rangle_{\!\psi_t}^2 
~\ttofor{\beta\rightarrow0}~ \myfrac{1}{2\beta} . 
\end{eqnarray}
Last, for the covariances, without the need of shifting the integration variables one can write 
\begin{eqnarray}
&&\overline{\overline{a_{ip}a_{iq}}} ~\ttofor{\beta\rightarrow0} 
\compint\diff\bm a_i \, a_{ip}\,a_{iq}\,{\mathscr P}(\bm a_i)  -  \overline{a_{ip}}\,\,\overline{a_{iq}}
= \compint\diff\bm a \, a_p\,a_q\,{\mathscr P}(\bm a)  -  \overline{a_{p}}\,\,\overline{a_{q}} \nonumber \\
&&= {\sum}_k {\cal P}_{\!\psi_t}\!(\bm\alpha_k)  \,\alpha_{kp} \,\alpha_{kq}
-  \overline{a_{p}}\,\,\overline{a_{q}} 
= \langle\hat A_p\, \hat A_q\rangle_{\!\psi_t} 
- \langle\hat A_p\rangle_{\!\psi_t}\langle\hat A_q\rangle_{\!\psi_t} . 
\end{eqnarray}
Let us define the set of stochastic variables 
\begin{equation}
\label{eq:db}
\diff\bm B = \sqrt{\myfrac{2\,\beta}{\mu}\,} 
\,{\sum_{i=1}^n} \big(\bm a_i - \langle\hat{\bm A}\rangle_{\!\psi_t}\big)   . 
\end{equation}
Taking the limit $\beta \to 0$ according to the prescription (\ref{eq:betamu}), $\mu $ and $n$ go to infinity 
in the same way, the conditions of the central limit theorem are satisfied so that the variables $\diff\bm B$ 
are Gaussian with the properties 
\begin{eqnarray}
\label{eq:dbstatprop}
&&\overline{\diff\bm B} = 0 , \nonumber \\
&&\overline{\diff B_p^2} = \myfrac{\,2\,\beta\,}{\mu} \myfrac{n}{\,2\,\beta\,}
= \myfrac{n}{\mu} = n\,\tau = \diff t , \nonumber \\
&&\overline{\diff B_p \,\diff B_q} 
= \myfrac{\,2\,\beta\,}{\mu} \,n 
\big(\!\langle\hat A_p\, \hat A_q\rangle_{\!\psi_t} 
- \langle\hat A_p\rangle_{\!\psi_t}\langle\hat A_q\rangle_{\!\psi_t} \!\big) \nonumber \\
&& \hskip 40pt = 2\,\beta \,\diff t 
\big(\!\langle\hat A_p\, \hat A_q\rangle_{\!\psi_t} 
- \langle\hat A_p\rangle_{\!\psi_t}\langle\hat A_q\rangle_{\!\psi_t}\!\big) 
~\ttofor{\beta\rightarrow0}~ 0 . 
\end{eqnarray}
Inserting the definition (\ref{eq:db}) into equation (\ref{eq:statechi}) one gets 
\begin{equation}
\ket{\chi_{t+dt}}  
= F \exp \big\{ \!-\gamma\big( \hat{\bm A}^2 - 2\,\hat{\bm A} \cdot \langle\hat{\bm A}\rangle_{\!\psi_t}\big) \diff t
+ \sqrt{\gamma}\,\hat{\bm A} \cdot \diff \bm B\,\big\} \ket{\psi_t} . 
\end{equation}
By expanding the exponential and using the rules of It\^o calculus one eventually obtains 
\begin{eqnarray}
&& \ket{\chi_{t+dt}}  = F \big[ 1 - {\textstyle\myfrac{1}{2}} \gamma\big(\hat{\bm A}^2 
- 4\,\hat{\bm A} \cdot \langle\hat{\bm A}\rangle_{\!\psi_t}\big) \diff t 
+ \sqrt{\gamma}\,\hat{\bm A} \cdot \diff \bm B\,\big] \ket{\psi_t} , \nonumber \\
&&\norm{\chi_{t+dt}}^{-\!1}  = F^{-\!1} \big[ 1 
- {\textstyle\myfrac{1}{2}} \gamma \langle\hat{\bm A}\rangle_{\!\psi_t}^2 \diff t 
- \sqrt{\gamma} \,\langle\hat{\bm A}\rangle_{\!\psi_t}\!\cdot\diff\tenbi B \big] , \nonumber
\end{eqnarray}
so that 
\begin{eqnarray}
\label{eq:stocproc}
 \diff\ket{\psi_t} &&= \ket{\psi_{t+\diff t}} - \ket{\psi_t} 
= \norm{\phi_{t+dt}} ^{-\!1}\,\ket{\phi_{t+dt}} - \ket{\psi_t}    \nonumber \\ \noalign{\vskip 9pt}
&&= \Big[
\sqrt{\gamma} \,\big(\hat{\bm  A} - \langle\hat{\bm A}\rangle_{\!\psi_t}\big) \cdot \diff\bm B 
- {\textstyle\myfrac{1}{2}} \gamma \big(\hat{\bm  A} - \langle\hat{\bm A}\rangle_{\!\psi_t}\big)^2 \diff t 
\,\Big]\, \ket{\psi_t} . 
\end{eqnarray}

By assuming that both the Schr\"odinger evolution and the stochastic process are there, and taking into account that 
the two terms in the stochastic differential equation~(\ref{eq:stocproc}) are of the order $\sqrt{\diff t}$ and $\diff t$, respectively, 
one can write on the whole 
\begin{equation}
\label{eq:contprocH}
 \diff\ket{\psi_t} = \Big[ - \myfrac{\rm i}{\hbar} {\hat H} \diff t +  
\sqrt{\gamma} \,\big(\hat{\bm  A} - \langle\hat{\bm A}\rangle_{\!\psi_t}\big) \cdot \diff\bm B 
- {\textstyle\myfrac{1}{2}} \gamma \big(\hat{\bm  A} - \langle\hat{\bm A}\rangle_{\!\psi_t}\big)^2 \diff t 
\,\Big]\, \ket{\psi_t} . 
\end{equation}
This is the form of the evolution equation normally assumed for continuous stochastic processes in Hilbert space, 
corresponding to eq.~(\ref{contproc}) with the addition of the term describing the Schr\"odinger dynamics. 

The above argument is worked out with reference to a case in which the quantity label runs over a finite numerable set. There are relevant situations in which the quantity label runs over a measurable continuous set. Two such cases will be examined in the following section, together with all the necessary changes.

\section{Three relevant implementations}
\label{sect:tri}

In the present Section we present  three physically most relevant implementations of discontinuous stochastic processes and the corresponding continuous evolution equations. 

As discussed in Section~\ref{sect:ifl}, both the discontinuous and the equivalent continuous processes are  characterized 
by the choice of the sharpened quantities $\hat A_p$, $p \in \{1, \dots, K \}$. 

The discontinuous process is further specified by a sharpening frequency $\mu$ and a sharpening accuracy $\beta$. The probability distribution of the hitting centres $a_{p,i}$ for the $i$-th hitting is assumed to be 
 \begin{equation}
{\mathscr P} \left( \psi_t \vert \bm a_i  \right) = \left( {\myfrac{\beta}{\pi}} \right)^{K/2} \! \!
\bra{\psi_t} \exp \left[ - \beta \sum_{p=1}^K \left(  {\hat A}_p - a_{p,i} \right)^2 \right] \ket{\psi_t} .
\end{equation}
 
The continuous process is ruled by equation~(\ref{eq:contprocH}) specified by the strength parameter $\gamma$ and by the properties of the Gaussian random variables 
\begin{eqnarray}
\label{eq:dbstatpropfin}
&&\overline{\diff\bm B} = 0 , \nonumber \\
&&\overline{\diff B_p^2} =  \diff t , \nonumber \\
&&\overline{\diff B_p \,\diff B_q} =  0 . 
\end{eqnarray}
For  equivalence of the two processes, the  parameter $\gamma$ must be given by $\gamma = \beta \mu / 2$. 

\vskip 12pt
\noindent 
$\bm{Distinguishable~particles}$

\vskip 3pt
For $N$ distinguishable particles the sharpened quantities are the three--dimensional positions   $\hat {\bm x}_l$,  
$l \in $ \{1, \dots, N\}. 

The discontinuous process~(\cite{Ghirardi:1985mt}) is defined by the localization frequency $\lambda_l$ for particle $l$ and by the localization accuracy 
$\alpha$. 
The probability distribution of localization centres $\overline{\bm x}_{l,i} $ for the $i$-th hitting on particle $l$ is 
 \begin{equation}
{\mathscr P} \left( \psi_t \vert  \overline{\bm x}_{l,i}  \right) = \left( {\myfrac{\beta}{\pi}} \right)^{\!  3/2} \! \!
\bra{\psi_t} \exp \left[ - \alpha  \left(  {\hat {\bm x}}_l - \overline{\bm x}_{l,i} \right)^2 \right] \ket{\psi_t} .
\end{equation}

The  corresponding continuous process is ruled by the stochastic differential equation 
\begin{eqnarray}
\label{eq:contprocHx}
&&\diff\ket{\psi_t} = \Big[ - \myfrac{\rm i}{\hbar} {\hat H} \diff t  \nonumber \\
 &+&   \, \sum_{l=1}^N \sqrt{\gamma_l} \big(\hat{\bm  x}_l - \langle\hat{\bm x}_l \rangle_{\!\psi_t}\big) \cdot \diff\bm B_l 
- {\textstyle\myfrac{1}{2}}  \sum_{l=1}^N \gamma_l \big(\hat{\bm  x}_l - \langle\hat{\bm x}_l \rangle_{\!\psi_t}\big)^2 \diff t 
\,\Big]\, \ket{\psi_t} , 
\end{eqnarray}
where the stochastic variables $\diff\bm B_l $ are $N$ independent three-dimensional Gaussian variables whose statistical properties are described in Eqs.~(\ref{eq:dbstatpropfin}). For  equivalence, the strength parameters $\gamma_l$ must be given by 
 $\gamma_l = \alpha \lambda_l / 2$.

\vskip 12pt
\noindent 
$\bm{Identical~particles}$

\vskip 3pt
In this case the localization effect is obtained by sharpening the particle density $\hat{N}({\bm x})$ around each point 
$\bm x $ in physical space.\footnote{J.S. Bell, private comunication, 1987. }
The particle densities can be defined in the second quantization language as 
\begin{equation}
\hat N({\bm x}) = 
\Big( \myfrac{\alpha}{2\pi} \Big)^{\! 3 /2}
\sum_s \int \! {\rm d} {\bm x}' \, 
\exp \left( - {\textstyle{1\over2}} \alpha \, \left({\bm x}' - {\bm x}\right)^2 \right) 
a^\dagger ({\bm x}' , s) a ({\bm x}' , s) , 
\end{equation}
$a^\dagger ({\bm x} , s)$ and $ a ({\bm x} , s)$ being the creation and annihilation operators of a particle at point $\bm x$ with spin component $s$. The {\it smooth} volume used to define the particle density has linear dimensions of the order of $1 / \sqrt{\alpha}$. 

For the discontinuous process the sharpening frequency and the sharpening accuracy of the density $\hat{N}({\bm x})$ are $\mu$ and $\beta$, respectively. It is to be noted that, because of the nature of the domain of the quantity label $\bm x$, the ``centre'' of the sharpening for the $i$-th hitting is now a  number density profile $n_i(\bm x)$ and its  probability density (in the functional space of number density profiles) is given by  
\begin{equation}
{\mathscr P} [n_i] = \vert C \vert^2 \bra{\psi_t} \exp\left[ 
- \beta \! \! \int  \! \! \diff {\bm x} \left(  \hat N({\bm x}) - n_i(\bm x)   \right)^2  \right] \ket{\psi_t} . 
\end{equation}
The coefficient $C$ is given by the normalization condition 
\begin{equation}
\int \! {\mathscr D} n \, {\mathscr P} [n] = 1. 
\end{equation}

The corresponding continuous process is ruled by the equation (\cite{Pearle:1988uh} and \cite{Ghirardi:1989cn})
\begin{eqnarray}
\label{eq:contprocHNx}
\diff\ket{\psi_t} &=& \Big[ - \myfrac{\rm i}{\hbar} {\hat H} \diff t 
 +  \sqrt{\gamma} \! \int \! \diff {\bm x} \left( \hat{N} (\bm  x) - \langle \hat{N} (\bm  x) \rangle_{\!\psi_t} \right)  \diff B ({\bm x})  \nonumber \\
&-& {\textstyle\myfrac{1}{2}} \gamma \! \int \! \diff{\bm x} \, \big( \hat{N} (\bm  x) - \langle \hat{N} (\bm  x) \rangle_{\!\psi_t}\big)^2 \diff t 
\,\Big]\, \ket{\psi_t} , 
\end{eqnarray}
where the Gaussian random variables $ \diff B ({\bm x})$  have the properties 
\begin{equation}
\overline{\diff B ({\bm x})} = 0,  \qquad \qquad \overline{\diff B ({\bm x}) \diff B ({\bm x}')} = \delta( \bm x - \bm x') \diff t . 
\end{equation}
For equivalence, the strength parameter $\gamma$ must be given by $\gamma = \beta  \mu / 2$. 

\vskip 12pt
\noindent 
$\bm{Several~kinds~of~identical~particles}$
\vskip 3pt
In the case of several kinds of identical particles, the most established formulation sharpens the mass density around each point in physical space by using a universal stochastic field $\diff B({\bm x})$.  The particle density operators $\hat{N}({\bm x})$ are then replaced, both in the discontinuous and the continuous processes, by the mass densities $\hat{M}({\bm x})$ where 
 \begin{equation} 
\hat{M}({\bm x}) = \sum_k m_k \hat{N}_k ({\bm x}) , 
\end{equation}
$m_k $ being the mass of the particle of kind $k$. 

The  continuous process (\cite{GBB:1995}) is ruled by Equation~(\ref{eq:contprocHNx}), with $\hat{N}({\bm x})$  replaced by $\hat{M}({\bm x})$. 

\section{Final considerations}

Some final comments are in order. From the discussion of Sections~\ref{sect:ifl} and \ref{sect:tri} it is apparent that 
the discontinuous processes bear the same generality as the continuous ones as far as their applicability to physical 
systems is concerned. In particular, contrary to what has been sometimes stated in the literature, discontinuous processes can be 
formulated for systems of identical particles or of several kinds of identical particles. To deal with such physical systems resorting to continuous formulations is not necessary. 

As explicitly shown in Section~\ref{sect:ifl}, discontinuous processes give rise, in a proper infinite frequency limit, to corresponding 
continuous ones, thus showing the  physical equivalence of the two formulations for sufficiently high hitting frequencies. Stated differently,  for any continuous process there is a discrete process which turns out to induce a dynamics as near as wanted to the corresponding continuous one and which becomes identical to it when the infinite frequency limit is taken. 

One could ask what really means ``sufficiently high'' frequencies.  From the purely formal point of view, the equivalence of 
discontinuous and continuous processes requires that, in the time interval $(t, t + \diff t ]$ one has a large enough number of hittings, so that the central limit theorem can be applied. Having said that, the effectiveness of the hitting process depends on the product $\beta \mu$, so that for fixed effectiveness one can still maintain a finite frequency, provided that a sufficient number of hittings occur on the time scale relevant to the solution of the measurement process. For the sake of simplicity, in Section~\ref{sect:ifl} it has been assumed that the hittings occur at evenly spaced times; in order to preserve time translation invariance (\cite{Ghirardi:1985mt}) one should use random times with a certain mean frequency. Then the mean frequency  has to be sufficiently large to guarantee that reduction takes place in the time interval of interest. There remains, however, a small probability that no reduction takes place. The same thing happens in the continuous process that leads certainly to a common eigenstate of the considered quantities only when $t \to \infty$. 

As a last comment we stress that the continuous processes, on one hand, are undoubtedly mathematically more elegant, while, on the other hand, the discontinuous processes show immediately the physical effect of reduction, so that, taken for granted the infinite frequency limit, they show the reduction properties of the continuous ones too. 

What really matters is the choice of the quantities induced to have a sharp distribution. We think it is important to stress the role 
of positions as the quantities that allow the strengthening of the process in going from microscopic to macroscopic degrees of freedom. 
In the case of distinguishable particles the variables undergoing the process are directly the positions of individual particles. In the case of identical particles or several kinds of identical particles the variables undergoing the process are the number or mass densities around the  running point  in physical space, that play the role of positions respecting the identity of particles.  
The final effect is again to make definite the position in space of macroscopic objects, thus providing a viable and conceptually simple  solution to the measurement problem. 

\bibliographystyle{plain}
\bibliography{HSSP}\label{refs}

\end{document}